\documentclass[prd,twocolumn,superscriptaddress,altaffilletter,showpacs,nofootinbib]{revtex4} 
\usepackage{amssymb,amsmath}
\usepackage{comment}
\usepackage{graphicx}

\begin{document}

\title{Novel echoes from black holes in conformal Weyl gravity}
\author{Mehrab Momennia}
\email{momennia1988@gmail.com}
\affiliation{Instituto de F\'{\i}sica y Matem\'{a}ticas, Universidad Michoacana de San
Nicol\'as de Hidalgo,\\
Edificio C--3, Ciudad Universitaria, CP 58040, Morelia, Michoac\'{a}n,
Mexico}
\date{\today }

\begin{abstract}
We reveal a novel class of echoes from black holes in conformal Weyl gravity
and show that they are generated due to the large-scale structure of the
cosmos, rather than near-horizon modifications of black holes as well as
wormhole spacetimes. To this end, we take into account the evolution of a
massive scalar perturbation on the background geometry of conformal Weyl
black holes and show that the corresponding effective potential enjoys a
double-peak barrier against the incident scalar waves. We perform the
calculations for the time evolution profiles of scalar perturbations to
understand how the linear term in the metric function and the cosmological constant
produce echoes. The Prony method is also employed to calculate the
quasinormal frequencies of the early-stage quasinormal ringing phase.
\end{abstract}

\maketitle

\section{Introduction}

After the merger phase of the black hole binaries, the newly formed remnant
object settles down to form a new black hole\footnote{The formation of black holes could also happen
in other channels, such as binary neutron stars and black hole-neutron star
binaries under certain conditions.} \cite%
{AbbottBBHmerger,AbbottBBHcoalescence}. This new deformed
black hole undergoes the ringdown stage \cite%
{VishveshwaraNature,PressRingdown,ChandrasekharDetweiler} and emits
gravitational waves in the form of quasinormal radiation that could be
detected through the gravitational wave interferometers \cite{Isi,MaSunChen}%
. Although there is a debate in the literature regarding the existence of
the first overtone in the ringdown signal of GW150914~\cite{BertiRingdown,CommentGW150914,ReplyGW150914},
such modes would be detectable through space-based gravitational wave
detectors in the future \cite{GWspectroscopy,RoadMap}. This final signal of
merger events is a superposition of damped oscillations with discrete
frequencies and decay times that are known as quasinormal modes (QNMs) [see 
\cite{Nollert,Kokkotas,BertiR,KonoplyaR} for review papers on QNMs].

In addition to interest in the QNMs from the gravitational wave astronomy
point of view, they attracted much attention in different branches of
fundamental physics as well. The QNMs spectrum governs the dynamic stability
of black holes undergoing the perturbations of various test fields \cite%
{InstabilityKZ,InstabilityWang,MomenniaLQBH}, the overdamped QN frequencies
play a crucial role in the semi-classical approach to quantum gravity \cite%
{Hod} and they have been used to fix the Barbero-Immirzi parameter in loop
quantum gravity \cite{Dreyer}. Besides, the eikonal QN frequencies correspond to
unstable circular null geodesics that are related to the black hole shadow
size \cite{CardosoUCNG,KonoplyaUCNG,MomenniaPRD,FurtherClarification} and
the QN modes in anti-de Sitter (adS) background refer to the decay of
perturbations of corresponding thermal state in the conformal field theory 
\cite{Horowitz,LemosAdS,KokkotasAdS,MehrabJHEP,MomenniaPLB}.

Recently, the QN modes of black holes undergoing various test field perturbations have been
investigated in modified theories of gravity, such as higher dimensional
Einstein gravity coupled to Yang-Mills field \cite{YangMills},
Einstein-Born-Infeld gravity \cite{BornInfeld}, loop quantum
gravity \cite{MomenniaLQBH,QuantumCorrectedBH,AxialLQG,LQBHjcap}, quantum corrected black hole metrics \cite{GongQuantum,CaoQuantum,DubinskyQuantum,MalikQuantum}, deformed
Schwarzschild black holes \cite{DeformedSchw}, spinning C-metric \cite%
{Cmetric}, and dirty black holes \cite{Dirty}. In addition, the QNMs of
black holes in Einstein-Gauss-Bonnet-adS spacetime \cite{EGB}, dark matter
halo \cite{DMH}, and $f\left( R\right) $\ gravity \cite{fRgravity} have been
studied\ and the anomalous decay rate of QNMs in Reissner-Nordstr\"{o}m dS
background has been examined \cite{YerkoAnomalous}. Furthermore, the QNMs of Dilaton-dS black holes \cite{DubinskyDilaton}, regular black holes \cite{Skvortsova,Skvortsova2025}, and Kaluza–Klein black holes \cite{KaluzaPLB,KaluzaMomennia,KaluzaPRD} have been explored.

Black hole perturbations in conformal Weyl gravity and corresponding QN modes have been investigated in a few articles so far \cite{MomenniaPRD,MomenniaPLB,MomenniaEPJC,KonoplyaWeyl,YerkoWeyl,Malik,Konoplya2025}. The QNMs have been studied for perturbations in the scalar \cite{MomenniaPRD,MomenniaPLB,YerkoWeyl, Konoplya2025}, electromagnetic \cite{MomenniaEPJC,Konoplya2025}, gravitational \cite{MomenniaEPJC}, and Dirac fields \cite{Malik,Konoplya2025}. It has been already shown that the quasinormal ringing of conformal black holes undergoing massless scalar perturbations contains dark matter and dS phases \cite{KonoplyaWeyl}. In this article, we shall show that for the massive test scalar field perturbations, a series of echoes dominates the ringdown profile of the modes. 

Although the early-stage quasinormal oscillations of black hole mimickers
are indistinguishable from those of the Schwarzschild ringdown, the signal
could be modified at later times by a series of echoes. The echoes have been
reported for QN modes of wormholes in \cite{WormholesEcho} for the first
time, and then studied for a large number of spacetimes including black hole
toy models (for instance, see an incomplete list \cite%
{CardosoEcho,NatureEcho,MaselliEcho,TsangEcho,WangEcho,KonoplyaEcho,YangEcho,GuoEcho,SangEcho}
and references therein). In this scenario, in addition to the photon sphere
peak in the effective potential of the wave-like master equation, a second
peak appears from various modifications in the background spacetime due to,
for example, quantum corrections or the free parameters of the gravitational
model under investigation\footnote{In the effective potential of some black hole toy models, more than two peaks have
been reported as well.}. Therefore, a substantial
component of the high-frequency initial signal crosses the potential barrier
and would be trapped between a double peak potential barrier, which leads to
a series of echoes detected by a distant observer.

Generally, the double-peak potentials and echoes appeared in limited cases,
like quantum-corrected black holes \cite{Abedi}, by taking into account
different equations of state for the ultra-compact objects \cite{StarEcho},
and considering matter distribution around the black hole \cite{BHshell}\
and wormhole \cite{wormshell}\ environment. In this paper, we show that this
second peak also appears in the case of conformal Weyl black holes due to
the linear term in the metric function, but on the right-hand side of the
angular momentum barrier. Even though this is also the case when one adds the
matter distribution around compact objects \cite{BHshell,wormshell}, we can
observe more intense echoes for our black hole case study due to large-scale structure of the Universe as well as the free
parameters of the theory purely arising as integration constants.

On one hand, there are still large uncertainties in measuring the background
test fields and the black hole parameters, such as angular momentum and
electric charge, by taking into account recent electromagnetic and
gravitational data from black holes \cite%
{AbbottBBHmerger,AbbottBBHcoalescence,M87,Sgr}. This uncertainty in
estimating parameters provides an opportunity to examine the strong field
regime and allows the existence of black hole models beyond Einstein gravity or general relativity coupled to various matter sources (see the incomplete list \cite%
{HartleThorne,EMDA,LQBH,Herdeiro,TopologicalBH,GausBonBH,EYM,Bah,Rastgoo,MariaLima} and references therein for some black hole metrics).
On the other hand, the standard cosmological model of the $\Lambda$-cold
dark matter successfully explains the current epoch of the Universe. Therefore, one should take into account the contribution of dark energy and
dark matter in the context of black hole physics in order to construct a
scenario in consistency with the large-scale structure of the cosmos. In this regard, more recently it has been demonstrated that the Hubble law emerges from the frequency-shift considerations of test particles revolving the
rotating Kerr black hole in asymptotically dS spacetime \cite{MomenniaHubble}%
, confirming that adding a cosmological constant term to the Einstein field
equations is indeed inevitable \cite{Carroll}.

In this direction, one may note that the cosmological constant term arises naturally
as an integration constant in\ the black hole solutions of conformal Weyl
gravity \cite{Birkoff}\ and this theory of general relativity can explain
the galactic rotation curves without the need for any dark matter \cite%
{GRCurves,DM}. Hence, in this paper, we are interested in
investigating the black hole solutions in conformal Weyl gravity in order to incorporate the
information of the large-scale structure of the Universe into the black hole
solutions. Therefore, we take into account gravitational modeling beyond Einstein gravity to study modifications in the ringdown signal of the black holes due to the free parameters of the theory.

The Lagrangian density of Weyl gravity is defined by the
square of the Weyl tensor and it is one of the successful and interesting
theories in higher derivative gravity scenario \cite{Weyl}. This theory of
gravity is invariant under local scale transformation of the metric $g_{\mu
	\nu }(x)\rightarrow \Omega ^{2}(x)g_{\mu \nu }(x)$, and therefore it is
unique up to the choice of the matter source for preserving the conformal
invariance property of spacetime.

Recently, different aspects of the conformal gravity have been recently studied, such
as self-dual gravitational instantons \cite{instantons}, black hole shadows \cite{shadow}, and its ghost problem \cite{ghost}. The role of the
linear term, appearing in the metric function of conformal Weyl black holes, has been explored
from the geometrical and astrophysical perspectives \cite{LRterm}. In
addition, the N\"{o}ther currents, black hole entropy universality, and
conformal field theory duality in conformal Weyl gravity have been
investigated \cite{CFTduality}.

This paper is organized as follows. The next section is devoted to a brief
review of black hole solutions in conformal Weyl gravity and the free
parameters of the spacetime. In Sec. \ref{Perturbations}, we consider the
perturbations of a test massive scalar field minimally coupled to the
gravitational theory and investigate criteria for having double-peak
potential barriers. Then, we briefly explain the time-domain integration and
the Prony method that are used to study the QN modes throughout the paper.
In Sec. \ref{Echoes}, the dependence of the early-stage quasinormal ringing
and subsequent echoes on the free parameter of conformal Weyl gravity is
investigated and the results are discussed. Finally, we finished our paper
with some concluding remarks through the last Sec. \ref{Conclusions}.

\section{Black hole solutions in conformal Weyl gravity}

In order to give a brief overview of conformal Weyl black holes, we start
with the unique conformally invariant action of Riemannian spacetime in four
dimensions as follows 
\begin{equation}
\mathcal{I}=\alpha \int_{\mathcal{M}}d^{4}x\sqrt{-g}C^{\mu \nu \rho \sigma
}C_{\mu \nu \rho \sigma },  \label{I3}
\end{equation}%
which $\alpha $ is a dimensionless parameter, $g$\ is the determinant of the
spacetime metric, and $C_{\mu \nu \rho \sigma }$ is the conformal Weyl
tensor with the following explicit form \cite{Weyl} 
\begin{eqnarray}
C_{\lambda \mu \nu \kappa } &=&R_{\lambda \mu \nu \kappa }+{\frac{1}{6}}%
R\left( g_{\lambda \nu }g_{\mu \kappa }-g_{\lambda \kappa }g_{\mu \nu
}\right)  \notag \\
&&-{\frac{1}{2}}\left( g_{\lambda \nu }R_{\mu \kappa }-g_{\lambda \kappa
}R_{\mu \nu }-g_{\mu \nu }R_{\lambda \kappa }+g_{\mu \kappa }R_{\lambda \nu
}\right) ,
\end{eqnarray}%
where $R_{\mu \nu \rho \sigma }$\ is the Riemann curvature tensor, $R_{\mu
\nu }$ is\ the Ricci tensor, and $R$ is the Ricci curvature scalar
corresponding to the background metric $g_{\mu \nu }$. One can show that the
Weyl tensor transforms covariantly as $C_{\mu \nu \rho \sigma }\left(
x\right) \rightarrow \Omega ^{-2}\left( x\right) C_{\mu \nu \rho \sigma
}\left( x\right) $ under conformal transformation,\ and thus the
gravitational action (\ref{I3}) remains invariant. We also note that Birkoff's theorem holds in the conformal
Weyl gravity \cite{Birkoff}.

By taking the variation of the action (\ref{I3}) with respect to the metric
tensor $g_{\mu \nu }$, one can obtain the field equations of the theory as
follows \cite{Bach} 
\begin{equation}
W_{\rho \sigma }=\left( \nabla ^{\mu }\nabla ^{\nu }+\frac{1}{2}R^{\mu \nu
}\right) C_{\rho \mu \nu \sigma }=0,  \label{GFE}
\end{equation}%
in which $W_{\mu \nu }$\ is the Bach tensor and $\nabla ^{\mu }$\ refers to
the standard\ covariant derivative compatible with the metric tensor $g_{\mu \nu }$%
. The Bach tensor has been linearized by Riegert and it was demonstrated
that there is a massless spin-$2$ ghost particle arising from the
fourth-order gravitational wave equations \cite{RiegertGhost}. Recently, a
couple of suggestions have been proposed to circumvent this ghost
instability in the linearized equations \cite{YangGW} as well as take
advantage of the conformal symmetry of Weyl spacetime in the metric \cite%
{MomenniaEPJC}.

\begin{figure*}[tbh]
\centering
\includegraphics[width=0.42\textwidth]{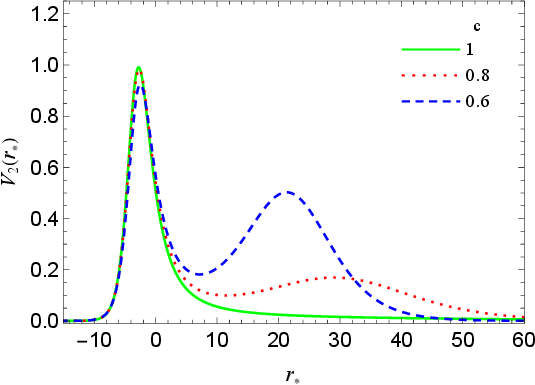} %
\includegraphics[width=0.4	\textwidth]{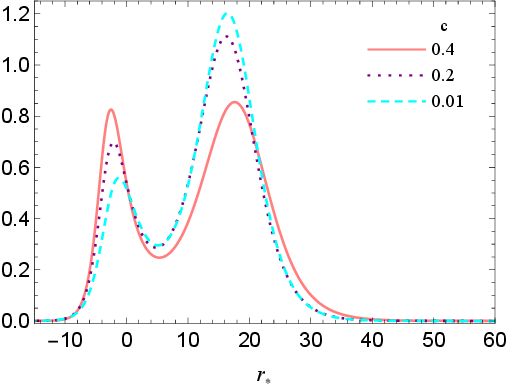}
\caption{The effective potential versus tortoise coordinate for $M=0.5$, $%
l=2 $, $\Lambda =0.001$, $\protect\mu =0.12$, and different values of the
constant parameter $c$. The continuous green curve denotes the Schwarzschild-dS
potential with a single photon sphere barrier. For $c<1$, the effective
potential forms an additional peak at some distances. For all cases, $%
V_{2}\left( r_{\ast }\right) $ vanishes at both infinities $r_{\ast
}\rightarrow \pm \infty $. The curves of $c\neq 1$ were shifted horizontally
for clarity.}
\label{Pot}
\end{figure*}

The static and spherically symmetric vacuum solutions of the Weyl gravity
describing black holes in this theory were first shown by Riegert \cite%
{Birkoff} and later obtained in \cite{MannheimKazanas}. It was shown that
the following spherically symmetric line element 
\begin{equation}
ds^{2}=-f(r)dt^{2}+f^{-1}(r)dr^{2}+r^{2}\left( d\theta ^{2}+\sin ^{2}\theta
d\varphi ^{2}\right) ,  \label{CS}
\end{equation}%
with the metric function 
\begin{equation}
f\left( r\right) =1-3\beta \gamma -\frac{2\beta -3\beta ^{2}\gamma }{r}%
+\gamma r-kr^{2},  \label{MF}
\end{equation}%
satisfies all components of the Bach tensor (\ref{GFE}) where $\beta $, $%
\gamma $, and $k$ are three integration constants. In order to compare our
results with the Schwarzschild black holes as well as in consistency with
the previous studies on the quasinormal modes of Weyl solutions \cite%
{MomenniaPRD,MomenniaEPJC,MomenniaPLB}, we found that it is convenient to
introduce the constants $\left\{ M,c,\Lambda \right\} $ in terms of the
integration constants\ $\left\{ \beta ,\gamma ,k\right\} $\ as below%
\begin{equation}
M=\frac{2\beta -3\beta ^{2}\gamma }{2},\ \ \ c=1-3\beta \gamma ,\ \ \
\Lambda =3k.
\end{equation}

By considering these new definitions for the integration constants, the
metric function (\ref{MF}) converts to%
\begin{equation}
f\left( r\right) =c-\frac{2M}{r}-\frac{c^{2}-1}{6M}r-\frac{\Lambda }{3}r^{2},
\label{mf}
\end{equation}%
that reduces to the standard\ Schwarzschild-dS solutions for $c=1$. In this
way, the constant parameter $c$ characterizes deviations from the Schwarzschild-dS black holes in a similar
manner as in \cite{MomenniaPRD,MomenniaEPJC,MomenniaPLB}. Therefore, we have
a spectrum of conformal solutions depending on the values of the free dimensionless
parameter\ $c$ obeying the condition $-1<c<2$\ in the nearly extreme regime 
\cite{MomenniaPRD}. Generally, the metric function (\ref{mf}) has at most
three real roots located at $r=r_{+}$ (the event horizon), $r=r_{Cosm}$ (the
cosmological horizon), and $r=r_{0}$ (a negative root) so that $%
r_{Cosm}>r_{+}$. In this paper, we shall study the perturbations in the
range $r_{+}\leq r\leq r_{Cosm}$.

It is notable to mention that in contrast with Einstein gravity that the
cosmological constant $\Lambda $ is added in the action by hand, it arises
purely as an integration constant in the Weyl gravity solutions (\ref{mf}).
Initially, it has been advocated that the linear $r$-term in the
metric function can explain the flat galaxy rotation curves without the need
for dark matter \cite{GRCurves,DM,MannheimKazanas} and there is still a debate in the literature regarding this matter \cite{Horne,Hobson,Mannheim2022}. But, here, we are going to show that
this term can be responsible for observing echoes in the ringdown signal of
black holes in Weyl gravity.

\section{Perturbation equations and time-domain integration \label%
{Perturbations}}

\subsection{Massive scalar field perturbations}

The wave equation and the effective potential of a massive scalar
perturbation minimally coupled to the background spacetime of the Weyl
solutions (\ref{mf}) are given by \cite{MomenniaPLB} 
\begin{equation}
\left( \frac{\partial ^{2}}{\partial t^{2}}-\frac{\partial ^{2}}{\partial
r_{\ast }^{2}}+V_{l}\left( r_{\ast }\right) \right) \Psi _{l}\left(
t,r_{\ast }\right) =0,  \label{WEq}
\end{equation}%
\begin{equation}
V_{l}\left( r_{\ast }\right) =f\left( r\right) \left( \mu ^{2}+\frac{l\left(
l+1\right) }{r^{2}}+\frac{f^{\prime }\left( r\right) }{r}\right) ,
\label{SP}
\end{equation}%
where $\mu $ is the test field mass, $l=0,1,2,...$\ are the multipole
numbers, and $r_{\ast }$ is the tortoise coordinate with the following
definition%
\begin{equation}
r_{\ast }=\int \frac{dr}{f\left( r\right) }.
\end{equation}

In this $r_{\ast }$ coordinate, the whole accessible space for the observer
lies between two infinities corresponding to the event horizon $r=r_{+}$ ($%
r_{\ast }\rightarrow -\infty $) and the cosmological horizon $r=r_{Cosm}$ ($%
r_{\ast }\rightarrow +\infty $).

\begin{figure}[tbp]
\centering
\includegraphics[width=0.45\textwidth]{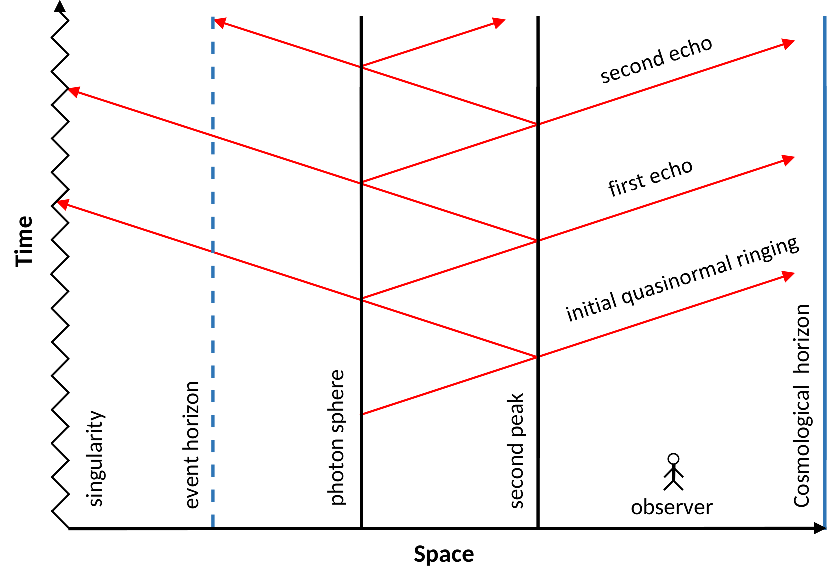}
\caption{Schematic Penrose diagram of the scalar wave echoes from the
conformal Weyl black hole solutions. A substantial component of the
high-frequency initial quasinormal ringing is trapped between the photon
sphere barrier and the secondary peak.}
\label{Penrose}
\end{figure}

The QN modes of conformal Weyl black holes are solutions to the wave
equation (\ref{WEq}) when satisfying the boundary conditions of purely
ingoing waves $\Psi \left( t,r_{\ast }\right) \sim e^{-i\omega \left(
t+r_{\ast }\right) }$ at the event horizon $r_{\ast }\rightarrow -\infty $
and purely outgoing waves $\Psi \left( t,r_{\ast }\right) \sim e^{-i\omega
\left( t-r_{\ast }\right) }$ at the cosmological horizon $r_{\ast
}\rightarrow +\infty $. Therefore, no waves are allowed to come from either
left or right infinity, and thus the distant observer only detects the
outgoing waves (see Fig. \ref{Penrose} and related discussion). Imposing these conditions on the wave equation (\ref{WEq})
lead to a discrete set of eigenvalues $\omega $\ with a real part
representing the actual oscillations of the test field and an imaginary part
giving its damping rate.

In order to see how the constant parameter $c$ produces the second peak in the
effective potential as well as affects the angular momentum peak, we
substitute the metric function (\ref{mf}) into (\ref{SP}) to get the explicit form of the potential in terms of the black hole parameters as below%
\begin{eqnarray}
V_{l}\left( r_{\ast }\right) &=&\left( \frac{c^{2}-1}{6M}\right) ^{2}-l(l+1)%
\frac{\Lambda }{3}+c\left( \mu ^{2}-\frac{2\Lambda }{3}\right)  \notag \\
&&-\frac{4M^{2}}{r^{4}}+\frac{2M\left[ c-l(l+1)\right] }{r^{3}}+\frac{cl(l+1)%
}{r^{2}}  \notag \\
&&-\frac{\left( c^{2}-1\right) \left[ c+l(l+1)\right] +4M^{2}\left( 3\mu
^{2}-\Lambda \right) }{6Mr}  \notag \\
&&-\frac{\left( c^{2}-1\right) \left( \mu ^{2}-\Lambda \right) }{6M}r-\frac{%
\Lambda }{3}\left( \mu ^{2}-\frac{2\Lambda }{3}\right) r^{2}.  \label{EffPot}
\end{eqnarray}

From this relation, we can see that the last two terms can produce the
second peak in the effective potential if we choose suitable values for the
parameters $\Lambda $, $c$, and $\mu $. We need the effective potential
vanishes at the cosmological horizon (that is also located on the right side of the second
peak), hence we impose the condition $\mu ^{2}>2\Lambda /3$\ on the
potential since the last $r^{2}$-term dominates for very large distances. On
the other hand, the only term that can raise the effective potential after the angular momentum peak (and before 
being suppressed by the last $r^{2}$-term) is the linear $r$-term which dominates at intermediate distances.
Therefore, by considering the linear term, in order for the effective potential grows and make the second
peak, one should impose the following restrictions%
\begin{equation}
\mu ^{2}>\Lambda ,\ \ \ c^{2}<1,  \label{Conditions}
\end{equation}%
on the parameter space. With these conditions at hand and for $l=0$, from
Eq. (\ref{EffPot}) we see that the combination of the set of terms $\left\{
r^{-4},r^{-3},r^{-1}\right\} $ raises the potential starting from zero value
at the event horizon and suppress it to form the first peak similar to the
standard Schwarzschild case, whereas the linear $r$-term raises the
effective\ potential again at intermediate distances and $r^{2}$-term
suppresses it at large distances to form the second peak.

The effective potential enjoys the second peak\ for $l>0$ as for the $l=0$
case and figure \ref{Pot} illustrates the behavior of the effective
potential (\ref{SP}) versus the tortoise coordinates for different values of
the constant parameter $c$. We find that this figure confirms our discussion
regarding the effective potential and shows a double-peak barrier in the
potential profile while the conditions (\ref{Conditions})\ are satisfied.

In this scenario, the role of the test field mass $\mu $\ is helping the $%
r^{2}$-term to suppress the effective\ potential and\ generate the second
peak. Otherwise, this term would be positive, and hence the potential would grow in the asymptotic region. Therefore, we see that the
first peak is produced by the black hole itself while the second peak is
produced due to the large-scale structure of the Universe characterized by
non-vanishing integration constants $c$\ (responsible for dark
matter) and $\Lambda $ (responsible for dark energy).

With this design at hand, we can observe echoes in the background spacetime
of Weyl solutions as well. In this case, the main burst is generated at the
photon sphere (left peak), but the frequency is not of the same order as the
Schwarzschild-dS black hole QNMs for $c\neq 1$. Then, a substantial
component of this high-frequency initial signal crosses the potential
barrier and would be trapped between the double peak barrier, which leads to
a series of echoes detected by a distant observer. These echoes occur in a
transient regime 
and each echo has a smaller amplitude and lower frequency
in comparison with previous ones.

The effects of the cosmological constant $\Lambda $ and the test scalar mass 
$\mu $ on the QNMs of black holes have been studied thoroughly so far. In
this study, we need to understand the dependence of the quasinormal ringing
on the constant parameter $c$ that characterizes deviations from the
Schwarzschild-$\Lambda $ black hole. Hence, we fix the free parameters as $%
M=0.5$, $l=2$, $\Lambda =0.001$, and $\mu =0.12$ in order to concentrate our
attention on the variation of $c$ in ringdown profiles.

\subsection{Ringing waveform}

We obtain the time-domain profile of modes\ by taking the integration of the
wave-like equation (\ref{WEq}). In order to obtain the time evolution of
modes, we follow the discretization scheme described in \cite{Pullin} and
rewrite the wave equation in terms of the light-cone coordinates $%
u=t-r_{\ast }$\ and $v=t+r_{\ast }$ as below%
\begin{equation}
\frac{\partial ^{2}\Psi _{l}\left( u,v\right) }{\partial u\partial v}=-\frac{%
1}{4}V_{l}\left( u,v\right) \Psi _{l}\left( u,v\right) ,  \label{WaveEq}
\end{equation}%
that takes into account the contribution of all the modes and
determines the behavior of
the asymptotic tails at late times. By applying the time evolution operator $%
e^{\Delta \partial _{t}}$ for small $\Delta $\ on $\Psi _{l}\left(
u,v\right) $\ and considering Eq. (\ref{WaveEq}), we get%
\begin{eqnarray}
&&\Psi _{l}\left( u+\Delta ,v+\Delta \right) \!\!=\Psi _{l}\left( u+\Delta
,v\right) +\Psi _{l}\left( u,v+\Delta \right)  \notag \\
&&-\Psi _{l}\left( u,v\right) -\frac{\Delta ^{2}}{8}\left[ V_{l}\left(
u+\Delta ,v\right) \Psi _{l}\left( u+\Delta ,v\right) \right.  \notag \\
&&\left. +V_{l}\left( u,v+\Delta \right) \Psi _{l}\left( u,v+\Delta \right) 
\right] ,
\end{eqnarray}%
where $\Delta $\ is the step size of the grids. This relation allows us to
obtain the values of $\Psi _{l}$ starting from the two null surfaces $%
u=u_{0} $\ and $v=v_{0}$ as initial data. In this paper, we consider $\Psi
_{l}\left( u,0\right) =1$\ at $v_{0}=0$,\ and use the Gaussian wave packet%
\begin{equation}
\Psi _{l}\left( 0,v\right) =\exp \left( -\frac{\left( v-v_{c}\right) ^{2}}{%
2\sigma ^{2}}\right) ,
\end{equation}%
centered on $v_{c}$ and having width $\sigma $ at $u_{0}=0$ as the initial
pulse. Then, we choose the observer to be located between the secondary peak and the cosmological
horizon corresponding to $r_{\ast }\sim 21$ to generate the quasinormal
ringing waveforms. Although the waveforms for various multipole numbers are
qualitatively similar, we choose the $l=2$ case in our analysis in order to
have a better illustration of the situation. This is because the $l=2$ case
shows the echoes after the ringdown phase more obvious in comparison with $%
l<2$.

\begin{figure}[tbp]
\centering
\includegraphics[width=0.45\textwidth]{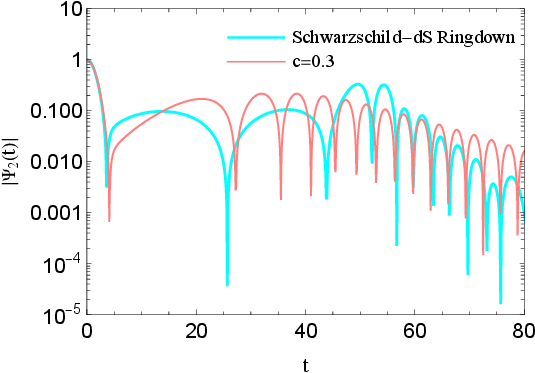}
\caption{The semilogarithmic plot of the time-domain profile for the scalar
field perturbations in the background of the standard Schwarzschild-dS
solutions ($c=1$) and the conformal Weyl black holes at the early-stage quasinormal ringing 
before echoes emerge. This figure shows that the initial quasinormal ringing of
conformal Weyl black holes does not follow the Schwarzschild-dS ringdown.}
\label{figC1}
\end{figure}
\begin{figure*}[tbh]
\centering
\includegraphics[width=0.4\textwidth]{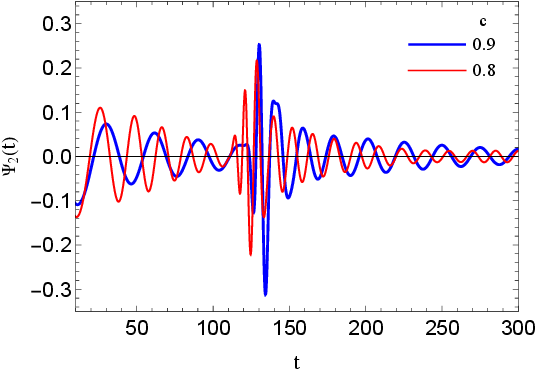} %
\includegraphics[width=0.4	\textwidth]{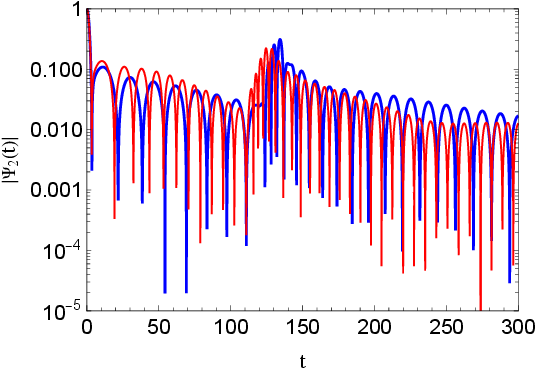}
\caption{The time evolution of the wave function $\Psi _{2}\left( t\right) $
(left panel) and its semilogarithmic plot (right panel) for the scalar field
perturbations of the conformal Weyl black holes. Both the real and imaginary
parts of the QN frequencies increase as the constant parameter $c$ decreases.}
\label{figC9}
\end{figure*}
\begin{figure*}[tbh]
\centering
\includegraphics[width=0.4\textwidth]{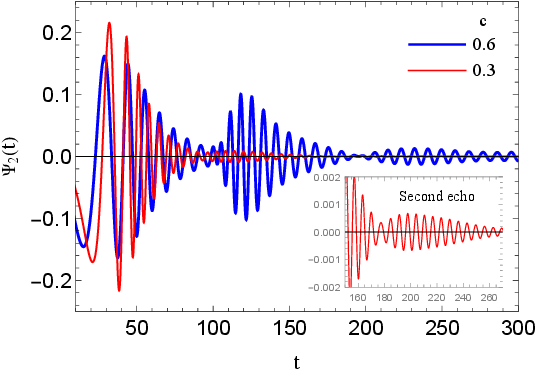} %
\includegraphics[width=0.4	\textwidth]{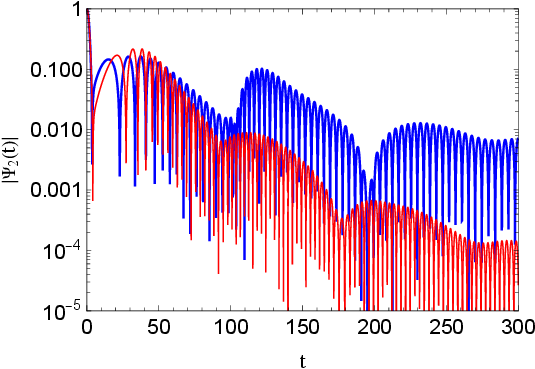}
\caption{The time-domain profile for the scalar field perturbations in the
background of the conformal Weyl black holes (left panel) and its
semilogarithmic plot (right panel). The perturbations live longer with lower
frequency as the constant parameter $c$ increases, and then a series of echoes
emerge at some time after the initial ringdown stage.}
\label{figC6}
\end{figure*}
\begin{center}
	\begin{table}[tbh]
		\scalebox{1} {\begin{tabular}{|c|c|c|c|}
				\hline\hline
				$c$ &  & $\omega \left( =\omega _{R}-i\omega _{I}\right) $ & $\mathcal{Q}$ \\ \hline
				$0.2$ &  & initial outburst, $1.051-0.0780i$, echoes & $6.7372$ \\ \hline
				$0.3$ &  & initial outburst, $0.9975-0.0740i$, echoes & $6.7399$ \\ \hline
				$0.4$ &  & initial outburst, $0.9220-0.0683i$, echoes & $6.7496$ \\ \hline
				$0.5$ &  & initial outburst, $0.8250-0.0610i$, echoes & $6.7623$ \\ \hline
				$0.6$ &  & initial outburst, $0.7068-0.0521i$, echoes & $6.7831$ \\ \hline
				$0.7$ &  & initial outburst, $0.5674-0.0413i$, echoes & $6.8693$ \\ \hline
				$0.8$ &  & $0.3859-0.0231i$, echoes & $8.3528$ \\ \hline
				$0.9$ &  & $0.2213-0.0123i$, echoes & $8.9959$ \\ \hline
				$1$ &  & initial outburst, $0.9686-0.1927i$ & $2.5132$ \\ \hline\hline
		\end{tabular}}
		\caption{The fundamental QNMs of the scalar field perturbations for the
			initial ringdown stage calculated by the Prony method and their
			corresponding quality factor $\mathcal{Q}=\omega _{R}/\left( 2\omega _{I}\right) $.
			The sixth-order Wentzel--Kramers--Brillouin expansion and improved
			Asymptotic Iteration Method have been also employed to obtain the QN
			frequencies of the Schwarzschild-dS black holes. To do so, we used the
			calculations presented in \cite{MomenniaPLB}\ for the special case $c=1$.
			The results are $\omega _{WKB}=0.9684-0.1926i$\ and $\omega
			_{AIM}=0.9684-0.1926i$\ which are in good agreement with the time-domain
			result.}
		\label{tab}
	\end{table}
\end{center}

\subsection{The Prony method}

In order to find the QN frequencies of the ringdown before the echo stage
emerges, we follow the procedure described in~\cite{Marple,PronyBerti}  and apply the Prony method to the generated\ signal. This method can
be employed for mining information from (damped) sinusoidal signals. Here, the main idea is to express the ringing waveform $%
\Psi _{l}\left( t\right) $ as a superposition of $p$ complex exponentials
with arbitrary amplitudes $A_{j}$ and phases $\phi _{j}$ as follows%
\begin{equation}
\Psi \left( t\right) =\sum_{j=1}^{p}A_{j}e^{-i\omega _{j}t+i\phi _{j}},
\end{equation}%
where $\omega _{j}=\omega _{R}-i\omega _{I}$\ are the QN frequencies with
the real part $\omega _{R}$ and the imaginary part $\omega _{I}$. To utilize
the Prony method, and since the time is discretized by $\Delta $, it is
convenient to rewrite this expression in the following slightly different
form~\cite{PronyBerti}
\begin{equation}
\psi _{k}\left( k\Delta \right) =\sum_{j=1}^{p}b_{j}z_{j}^{k},
\label{NewModeExpansion}
\end{equation}%
where $\psi _{k}$\ is the value of $\Psi \left( t\right) $ at the $k$th step
and the new unknown complex parameters are given by%
\begin{equation}
z_{j}=e^{-i\omega _{j}\Delta },  \label{QNMrelation}
\end{equation}%
\begin{equation}
b_{j}=A_{j}e^{i\phi _{j}}.
\end{equation}

Indeed, this method allows us to obtain $z_{j}$'s, and since $\psi _{k}$\
and $\Delta $\ are known,\ we can calculate the QN frequencies from Eq. (\ref%
{QNMrelation}) as below 
\begin{equation}
\omega _{j}=\frac{i}{\Delta }\ln \left( z_{j}\right) .  \label{qnf}
\end{equation}

In order to find $z_{j}$'s, as the first step, one can introduce $k\rightarrow k-m$ in Eq. (%
\ref{NewModeExpansion}) to
obtain a new summation as follows 
\begin{eqnarray}
\sum_{m=0}^{p}\left( \psi _{k-m}\right) \alpha _{m} &=&\sum_{m=0}^{p}\left(
\sum_{j=1}^{p}b_{j}z_{j}^{k-m}\right) \alpha _{m}  \notag \\
&=&\left( \sum_{j=1}^{p}b_{j}z_{j}^{k-p}\right) \sum_{m=0}^{p}\alpha
_{m}z_{j}^{p-m}.  \label{ZeroSum}
\end{eqnarray}

Now, by taking into account\ the following polynomial function%
\begin{equation}
\digamma \left( z\right) =\prod_{j=1}^{p}\left( z-z_{j}\right)
=\sum_{m=0}^{p}\alpha _{m}z^{p-m};\ \ \ \ \alpha _{0}=1,  \label{polyfun}
\end{equation}%
one finds that the second summation in the last equality of Eq. (\ref%
{ZeroSum}) is zero, hence from its left-hand side, we have%
\begin{equation}
\psi _{k}+\sum_{m=1}^{p}\alpha _{m}\psi _{k-m}=0,  \label{AlphaEq}
\end{equation}%
that can be used to find the series coefficients $\alpha _{m}$ presented in the
polynomial function (\ref{polyfun}). To do so, we assume the ringdown
waveform starts at $t_{i}=0$\ and finishes at some time $t_{f}=\left(
2p-1\right) \Delta $. Therefore, by substituting $k=p,p+1,...,2p-1$\ in Eq. (%
\ref{AlphaEq}), we can find $p$\ linear equations for $p$\ unknown series
coefficients $\alpha _{m}$. Once the coefficients $\alpha _{m}$ are found by solving these $p$ equations, 
we numerically solve the polynomial function (\ref{polyfun}) for the roots $%
z_{j}$. Finally, Eq. (\ref{qnf}) can be used to calculate the quasinormal
frequencies.

\section{Echoes from Weyl black hole solutions \label{Echoes}}

Here, we investigate the dependence of the quasinormal ringing in the
background of Weyl solutions on the constant parameter $c$ by employing the
time-domain integration approach described in the previous section. We
recall that the integration constant $c$ measures deviations from the
Schwarzschild-dS black holes for $c\neq 1$. The time-domain profile of the
modes is illustrated in Figs. \ref{figC1}-\ref{figC6} for various values of
the constant parameter $c$ while the other free parameters are fixed to $M=0.5$, 
$l=2$, $\Lambda =0.001$, and $\mu =0.12$. Generally, according to the time
evolution of the modes, we can observe that the echoes appear after the
period of initial quasinormal oscillations and dominate the signal.

Besides, the fundamental QN frequencies of the initial ringdown stage and
their corresponding quality factor have been calculated and the results are
given in Table \ref{tab}. However, note that the accuracy of quasinormal frequencies extracted
from the ringdown waveform data is sensitive to the temporal range of considered early-stage quasinormal ringing. Furthermore, from the quality
factor $\mathcal{Q}$ of the QN frequencies, we find that the signal of the conformal
Weyl black hole has higher quality compared to the Schwarzschild-dS one.
Hence, the Weyl solutions are better oscillators and their signal is more
likely to be detected.

Fig. \ref{figC1}\ illustrates a comparison between the initial quasinormal
ringdown of the conformal Weyl black holes and the Schwarzschild-dS
solutions before a series of echoes appear. This figure indicates that the initial
quasinormal ringing of conformal Weyl black holes does not follow the
standard\ Schwarzschild-dS ringdown. For the special case $c=0.3$ and from
this figure, we see that the real part of the QN frequencies for the Weyl
solutions is close to the Schwarzschild-dS black holes, but the
perturbations live longer in the former background (also see Table \ref{tab} for
quantified QN frequencies of this early-stage ringing phase and compare $%
c=0.3$\ row with $c=1$\ row).

However, the situation would change if the constant parameter $c$\ varies. As
one can see from Table \ref{tab} and Figs. \ref{figC9}-\ref{figC6},
both the real and imaginary parts of the QN frequencies decrease with an
increase in the constant parameter $c$. Hence, the perturbations live longer
with lower frequency as the constant parameter $c$ increases. Furthermore, these
figures show that a distinctive picture of echoes appears after the initial
oscillatory ringing and dominates the time evolution of the modes. This
behavior can be observed for the range $c\in \left( -1,1\right) $ of the
constant parameter $c$ subject to choosing suitable values for the rest of the free
parameters. These echoes disappear for $|c| \ge 1$.

Strictly speaking, one must obey the conditions (\ref{Conditions}) on the
set of parameters $\left\{ \mu ,c,\Lambda \right\} $ in order to observe
echoes raising immediately after the quasinormal ringdown period. But if we
violate one of these constraints on the free parameters, we would see that
quasinormal oscillations dominate the signal and echoes disappear. Besides,
in the cases studied here, we see that the effective potential is positive
definite\ and the perturbations decay with the time that guarantees the
dynamical stability of the\ spacetime undergoing massive scalar perturbations.

It is worthwhile to recall that, unlike the well-known echoes produced due
to near-horizon modifications, such as horizon-corrected black holes and
wormholes, from our analysis, we see that the echoes in conformal Weyl
spacetime are generated because of the large-scale structure of the
Universe. This new class of echoes emerges due to trapped modes between the
photon sphere peak and the second barrier generated by the dark matter and
dark energy contents of the cosmos, rather than trapped signals between the
angular momentum peak and the near-horizon area (see Fig. \ref{Penrose} for a
schematic illustration of the situation and compare with Fig. $1$\ of Ref. 
\cite{Abedi} for corrections near the horizon).

As the final remark, we should note that the large-scale terms presented in
the spacetime (\ref{CS})\ that are responsible for observing these echoes,
i.e., the linear $r$-term (responsible for dark matter) and the cosmological constant $r^{2}$-term (responsible for dark energy), arose
quite naturally as integration constants in the conformal Weyl gravity
theory. Therefore, these novel echoes emerged due to the large-scale structure of the universe and differ from the ones that appeared in black holes due to near-horizon modifications~\cite{Abedi}, the compact stars because of discontinuity in their effective potential~\cite{discontinuityPotential}, wormhole spacetimes~\cite{WormholesEcho}, and considering massive thin shells in the black hole~\cite{BHshell} and wormhole~\cite{wormshell} environment.

\section{Outlook and Conclusions \label{Conclusions}}

We have employed the time-domain integration in order to analyze the
quasinormal ringing of black hole solutions in conformal Weyl gravity and
used the Prony method to extract the quasinormal frequencies of the
early-stage oscillations. The metric under consideration depends on an
integration constant, which is denoted by $%
c$ throughout the paper, and characterizes deviations from the standard Schwarzschild-dS metric;
The conformal Weyl black holes deviate from the Schwarzschild-dS black holes
if the free parameter $c$ deviates from one, hence the former solutions
reduce to the latter ones provided $c=1$.

We have shown that the deviations from the Schwarzschild-dS black holes are
characterized by echoes such that the first stage signal is dominated by a
series of echoes at later times. The echoes could be observed for the range $%
c\in \left( -1,1\right) $ of the constant parameter $c$ for our black hole case
study. In addition, one may consider the possibility that if this class of echoes could be produced after the merger
phase and exist in nature, their contribution to the recent pulsar timing
array observations \cite{PTA,NANOGrav} might be inevitable. This is because
each echo that propagates through the cosmos has a smaller amplitude and 
\emph{lower frequency} in comparison with previous ones, hence finally,
leads to Nano-Hertz gravitational waves. However, this impact is not clear yet and it requires more investigations which we leave here as a proposal. 

We have seen that this phenomenon emerged due to the large-scale structure
of the Universe, unlike wormholes and horizon-corrected black holes, because
the echoes are induced due to the appearance of the second peak on the right
side of the photon sphere peak; This second peak basically appeared because
of the presence of the linear $r$-term (responsible for dark matter) and the quadratic $r^{2}$-term (responsible for dark energy) in the
metric function that encode the large-scale information of the cosmos and
arose quite naturally as integration constants in the conformal Weyl gravity
theory. Therefore, we have seen that the echoes emerge when one assigns conformal symmetry to the astrophysical black hole spacetimes. Furthermore, even after conformal symmetry breaking, the echoes are present if nature chooses the constant
parameter value in the interval $-1<c<1$.

\section{acknowledgments}

The author acknowledges SNII and was supported by the
National Council of Humanities, Sciences, and Technologies of Mexico
(CONAHCyT) through Estancias Posdoctorales por M\'{e}xico Convocatoria
2023(1) under the postdoctoral Grant No. 1242413.

\end{document}